\documentclass[12pt,eqsecnum,showpacs,preprintnumbers,amssymb,aps]{revtex4}
\usepackage{graphicx}
\usepackage{dcolumn}
\usepackage{bm}

\begin{document}

\title{Theoretical determination of lifetimes of metastable states in Sc III and Y III} 
\vspace{0.5cm}

\author{B. K. Sahoo \protect \footnote[2] {E-mail: bijaya@mpipks-dresden.mpg.de} \\
{\it  Max Planck Institute for the Physics of Complex Systems, N\"othnitzer Str. 38, D-01187 Dresden, Germany}}
\author{\\ H. S. Nataraj, B. P. Das and R. K. Chaudhuri \\
{\it Non-Accelerator Particle Physics Group, Indian Institute of
Astrophysics, Bangalore- 560 034, India }}
\author{\\ D. Mukherjee\\
{\it Department of Physical Chemistry, 
Indian Association for the Cultivation of Sciences, Kolkata-700 032, India}}
\date{\today}
\vskip1.0cm

\begin{abstract}
\noindent
Lifetimes of the first two metastable states in Sc$^{2+}$ and Y$^{2+}$
are determined using the relativistic coupled-cluster
theory.
There is a considerable interest in studying the electron correlation effects in these ions as 
though their electronic configurations are similar to the neutral alkali
atoms, their structures are very different from the latter.
We have made a comparative study of the correlation trends between the above doubly ionized 
systems with their corresponding neutral and singly ionized iso-electronic systems. The lifetimes of the excited states of these ions are very important in the field of astrophysics, especially for the study of post-main sequence evolution of the cool giant stars.
\end{abstract} 

\maketitle

\section{Introduction}
With the recent progress in high performance computational techniques, it is now possible to 
carry out accurate calculations on heavy atomic systems using powerful relativistic many-body theories \cite{sergey,rodrigues,bijaya2006}. It is not possible to treat all the heavy atomic systems by a single many-body theory due to the 
differences in their electronic structures. The relativistic coupled-cluster 
(RCC) theory has been successfully applied earlier to a wide range of atomic 
properties \cite{das} including the lifetime calculations of Pb$^+$ \cite{bijaya1} and alkaline-earth ions \cite{bijaya2}. The role of electron correlation 
effects is dramatic in some of those cases. In the present work, we have carried out the lifetime calculations of the lowest $d_{5/2}$ and $s_{1/2}$ states in Sc III and Y III ions. 

\begin{figure}[h]
\includegraphics[width=8.0cm,clip=true]{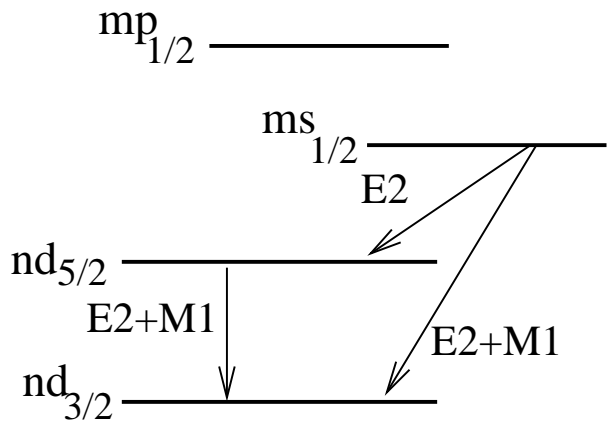}
\caption{Schematic diagram of the energy levels and transition lines between the $ms$ and $nd$ states in Sc$^{2+}$ and Y$^{2+}$. The $n$ and $m$ represent the principal quantum numbers of the ground and metastable states where $n=3$ and $m=4$ for Sc$^{2+}$ and $n=4$ and $m=5$ for Y$^{2+}$ .}
\label{fig1}
\end{figure}
Generally, the forbidden transition lines find numerous interesting applications in astrophysics which includes the determination of the elemental abundances in different celestial objects and also in the laboratory astrophysics and plasma physics studies. The spectral lines of many alkaline-earth ions including Y and its doubly charged ions have been observed in the spectra of many hot stars like Ap stars \cite{redfors} and these ions seem to be over abundant in Hg-Mn stars compared to their normal stellar abundances. This anomaly observed over abundances than that expected from the theoretical estimations based on the s-process of neutron capture is studied extensively in the literature \cite{adelman,guthrie}. Thus, the 
 study of adjacent elements of the same row of the periodic table like, the Sr-Y-Zr triad of elements, can be used to understand the nucleo-synthetic origin of these elements and it can reveal the post-main-sequence
evolution of the chemically peculiar stars, such as Hg-Mn stars \cite{redfors}. The Y III and Sc III ions can also be used for the diagnostics and modeling of the stellar plasma \cite{dimitri}. Recently, Brage et. al. \cite{brage} have used the multi-configuration Hartree-Fock (MCHF) method in calculating the lifetimes of the 5p, 5d and 6s states of Y III which were measured earlier by Maniak et. al. \cite{maniak}. However the lifetimes of the $4d_{5/2}$ and $5s_{1/2}$ states in Y III seem to be unreported so far. Though there are a large number of results available for the transition wavelengths of Sc III, it seems that there 
are no measurements on either the lifetimes or the transition probabilities of 
the $3d_{5/2}$ and $4s_{1/2}$ states. For this reason, we have carried out the 
calculations of the lifetimes of the lowest $d_{5/2}$ and $s_{1/2}$ metastable 
states in Sc III and Y III employing a powerful many-body theory known as the 
RCC theory. As the experimental wavelengths are  known to high accuracy we have used them in our calculations in order to minimize the errors in our lifetime estimations. 
The low-lying excited states and their electromagnetic decay channels are shown in the figure \ref{fig1}. Even though 
Sc III and Y III belong to K- and Rb- iso-electronic sequence, respectively, 
the ground states of the former two are $n\,d_{3/2}$ states ($n=3$ for Sc III and $n=4$ for Y III) unlike the latter 
two for which the ground states are $(n+1)\,s_{1/2}$. Again though the considered systems are the doubly charged ions, they have the same ground states as that of their neutral counterparts. Thus the ions considered in this work are interesting from the theoretical and as well as astrophysical point of view.
We analyze the role of correlation effects in these calculations
and compare their trends with the previously studied lifetimes of the $d_{5/2}$ states \cite{safronova} in
alkali atoms and alkaline earth metal ions \cite{bijaya2}. 

\section{Theory and method of calculation}
The transition probabilities for the spontaneous emissions due to E2 and M1 electromagnetic transitions from the state $i$ to the state $f$ are given by,
\begin{eqnarray}
A^{\text{E2}}_{i \rightarrow f} &=& \frac {1.11995 \times 10^{18} }{[J_i] \lambda^5} S^{\text{E2}} \label{eqn1} \\
A^{\text{M1}}_{i \rightarrow f} &=& \frac {2.69735 \times 10^{13} }{[J_i] \lambda^3} S^{\text{M1}},\label{eqn2}  
\end{eqnarray}
where, $[J_i]=2J_i+1$ is the degeneracy of the state $i$, $S = |\langle f |O| i \rangle|^2$ is the
transition line strength in atomic unit (au) for the corresponding transition operator $O$ and wavelength $\lambda$ (\AA).

As shown in the figure \ref{fig1}, the $d_{5/2}$ state can decay to the
$d_{3/2}$ state in the considered systems via E2 and M1 channels; hence the total transition probability for the 
$d_{5/2}$ state can be expressed as,
\begin{eqnarray}
A_{n d5/2} &=& A^{\text{M1}}_{n d5/2 \rightarrow n d3/2} + A^{\text{E2}}_{n d5/2 \rightarrow n d3/2}.
\label{eqn31}
\end{eqnarray}
Similarly, the $s_{1/2}$ state can decay to the $d_{3/2}$ state in these systems via E2 
and M1 channels and decay to the $d_{5/2}$ state via E2 channel; the resulting total 
transition probability is
\begin{eqnarray}
A_{m s1/2} &=& A^{\text{M1}}_{m s1/2 \rightarrow n d3/2} + A^{\text{E2}}_{m s1/2 \rightarrow n d3/2} + A^{\text{E2}}_{m s1/2 \rightarrow n d5/2}  .
\label{eqn32}
\end{eqnarray}
In the above expressions, $n$ and $m$ represent the principal quantum numbers 
of the $d$ and $s$ states; where, $n=3$ and $m=4$ for Sc III and $n=4$ and $m=5$  for Y III. The single particle reduced matrix elements of E2 and M1 operators
used in our calculations can be found from \cite{johnson}.
The lifetimes of these states are determined by taking the inverse
of their total probabilities. 

We use the following RCC
ansatz to construct the wave functions for one valence states ($|\Psi_v \rangle$) in the considered systems;
\begin{eqnarray}
|\Psi_v \rangle = e^T \{1+S_v\} |\Phi_v \rangle, 
\label{eqn12}
\end{eqnarray}
with the reference state $|\Phi_v \rangle= a_v^{\dagger}|\Phi_0\rangle$  
for the valence electron $v$ and $|\Phi_0\rangle$ is the Dirac-Fock (DF) state 
of the closed-shell system constructed by the Dirac-Coulomb (DC) Hamiltonian. 
This is the relativistic method of the universal valence coupled-cluster
theory proposed by Mukherjee et al \cite{mukherjee} and later was compactly formulated
by Lindgren \cite{lindgren1}. In the above RCC wave function expression,
we define $T$ and $S_v$ as the closed-shell and open-shell excitation operators
 which excite core electrons and valence with core electrons, respectively.
We have considered all single, double and a subset of important triple excitations approximation in our calculations
which is known as the CCSD(T) method in the literature \cite{bijaya2,bijaya3}. It has been found 
in many studies in the similar systems as considered here that the CCSD(T)
method is powerful enough to account the correlation effects to evaluate
various properties to high accuracy \cite{bijaya1,bijaya2,bijaya3,bijaya4}. This is primarily because of 
the fact that this approach considers the most important correlation effects
like core-polarization and pair-correlation effects to all-orders \cite{bijaya4}.

The transition matrix elements for the operator $O$ are calculated in this
approach using the following expression; 
\begin{eqnarray}
 \langle O \rangle_{fi}  &=& \frac {\langle\Psi_f | O | \Psi_i \rangle} { \sqrt{ \langle\Psi_f|\Psi_f\rangle} \sqrt{ \langle\Psi_i|\Psi_i\rangle} } \nonumber \\
 &=& \frac {\langle \Phi_f | \{1+S_f^{\dagger}\} e^{T^{\dagger}} O e^T \{1 + S_i\} | \Phi_i\rangle } { \sqrt{N_f} \sqrt{N_i}} \nonumber \\
 &=& \frac {\langle \Phi_f | \{1+S_f^{\dagger}\} \overline{O} \{1 + S_i\} | \Phi_i\rangle } { \sqrt{N_f} \sqrt{N_i}} , \label{eqn14}
\end{eqnarray}
where, $\overline O = e^{T^{\dagger}} O e^T$. Since the above expression contains many non-linear terms those will be computationally expensive if they
are calculated directly, we follow the extended Wick's theorem \cite{lindgren2} to break these
terms in simple form as discussed below. Although these procedures are given
in detail earlier \cite{bijaya1,bijaya4}, we mention them briefly here. First, we divide
$\overline O$
as the effective one-body, two-body etc. operators and later
these terms are sandwiched between the necessary $S_v$ operators. To save
the computational time, we calculate and save only the one-body terms which are 
later used where
other terms are computed directly in the final property calculations.
It has to be noticed that the fully contracted $\overline O$ terms do not
contribute in the present calculations. Contributions from the normalization
of the wave functions (Norm.) are accounted by,
\begin{eqnarray}
\text{Norm.} = \langle \Psi_f\, |\, O\, |\, \Psi_i \rangle \{ \frac {1}{\sqrt{N_f N_i}} - 1 \} ,
\end{eqnarray}
where, $$N_v = \langle \Phi_0 | e^{T^{\dagger}} e^T|\Phi_0 \rangle + \langle \Phi_v | \{ S_v^{\dagger} e^{T^{\dagger}} e^T S_v \}|\Phi_v \rangle$$ for the corresponding valence electrons $v\, (= i, f)$.

\section{Results and Discussions}
In Table \ref{tab1}, we present the wavelengths and transition line strengths of the low-lying transitions from the lowest $d_{5/2}$ and
$s_{1/2}$ states to the ground states of Sc III and Y III. The transition wavelengths for Sc III are taken 
from the critically evaluated atomic data base of National Institute for the Standard and Technology (NIST) \cite{nist} and the transition wavelengths for Y III are taken from \cite{epstein1975} for the given
transitions. The calculated transition strengths are used to determine the corresponding transition probabilities  which are given in the last column of the above mentioned table.

\begin{table}
\caption{Transition wavelengths, line strengths and transition probabilities ($A$s) of different low-lying transitions in Sc$^{2+}$ and Y$^{2+}$.}
\begin{ruledtabular}
\begin{center}
\begin{tabular}{lcccc}
Transition &  $\lambda$ & Transition & Strength & $A$\\
 ($f \rightarrow i$) & (in \AA) & channel & (au) & $sec^{-1}$\\
\hline\\
 &  & & &\\
\underline{Sc$^{2+}$} \\
$3d \ ^2D_{5/2}\rightarrow 3d \ ^2D_{3/2} $ & 505970 & M1 & 2.3751 &8.2434\,E-5 \\
 &  & E2 & 2.7164  & 1.5291\,E-11\\
 &  & &  \\
$4s \ ^2S_{1/2}\rightarrow 3d \ ^2D_{3/2} $ & 3916 & M1 & 7.9495E-7 & 1.786\,E-4\\
 &  & E2 & 12.9157 &7.8584 \\
$4s \ ^2S_{1/2}\rightarrow 3d \ ^2D_{5/2} $ & 3946 & E2 & 19.5007 &11.413\\
 &  & &  \\
\underline{Y$^{2+}$} \\
$4d \ ^2D_{5/2} \rightarrow 4d \ ^2D_{3/2}$ & 138093 & M1 & 2.3822& 4.0667\,E-3\\
 &  & E2 & 9.9573&3.7011\,E-8  \\
 &  & &  \\
$5s \ ^2S_{1/2}\rightarrow 4d \ ^2D_{3/2} $ & 13392 & M1 & 1.1136E-7 &6.2544\,E-7\\
 &  & E2 & 37.6554 &4.895\,E-2 \\
$5s \ ^2S_{1/2}\rightarrow 4d \ ^2D_{5/2} $ & 14830 & E2 & 57.7644 &4.5090\,E-2\\
\end{tabular}
\end{center}
\end{ruledtabular}
\label{tab1}
\end{table}
The calculated total transition probabilities for the lowest $d_{5/2}$ and $s_{1/2}$ states are given in Table \ref{tab2}. The lifetimes  are determined by taking
the reciprocal of these quantities for the corresponding states. The M1 
transition probabilities between the fine structure levels, i.e. the $d_{5/2}\rightarrow d_{3/2}$ transitions are 5-6 orders of magnitude larger than the E2 transitions, where as, the M1 transition probabilities for the $d_{3/2}\rightarrow s_{1/2}$ transitions are 5-6 orders smaller than the corresponding E2 transitions 
(cf. Table \ref{tab1}) for both the ions. Thus, the M1 transitions play crucial
roles in determining the lifetimes of the $d_{5/2}$ states but it is the other 
way around for the $s_{1/2}$ states as the E2 transitions play crucial roles in
their lifetime determinations. As it can be seen from Table \ref{tab2} that
the lifetime of the $3d_{5/2}$ state in Sc III is very large
implying that it is a highly forbidden state. 
However, the lifetime of the $4s_{1/2}$ state is a few
 fraction of a second. On the other hand, the lifetime of the $4d_{5/2}$ state  
in Y III is $246$ seconds which again is a long lived metastable state, where as, the $5s_{1/2}$ state has a lifetime of only $10$ seconds. This also indicates that the correlation effects in the calculations of M1 transition amplitudes of
the $d_{5/2} \rightarrow d_{3/2}$ transitions are very small and hence
these results can be ascribed as relativistic in origin. The transition probabilities of
the states given in Table \ref{tab2} will be very useful
in astrophysics for the accurate determination of the abundances of these elements. We estimate that our lifetime results can have an error bar of less than 2\%. The source of errors includes the truncation of the coupled-cluster series at the double excitations and partial inclusion of triple excitations,  the omission of Breit and QED effects in our calculations.
\begin{table}
\caption{The transition probabilities and the lifetimes of the $s_{1/2}$ and $d_{5/2}$ states in Sc$^{2+}$ and Y$^{2+}$.} 
\begin{ruledtabular}
\begin{center}
\begin{tabular}{lcc}
State & Transition Probability & Lifetime \\
 & (sec$^{-1}$) & (sec) \\
\hline\\
 &  & \\
\underline{Sc$^{2+}$} \\
$3d \ ^2D_{5/2}$ & 8.2434\,E-5 & 12130.86  \\
$4s \ ^2S_{1/2}$ & 19.2715 & 0.0519 \\
 &  &  \\
\underline{Y$^{2+}$} \\
$4d \ ^2D_{5/2}$ & 4.0668\,E-3  & 245.89 \\
$5s \ ^2S_{1/2}$ & 9.4041\,E-2 & 10.63 \\
\end{tabular}
\end{center}
\end{ruledtabular}
\label{tab2}
\end{table}

\begin{table}
\caption{Reduced matrix elements of the M1 and E2 transition operators from
$d_{5/2}$ states to the ground states in au in the considered systems.}
\begin{ruledtabular}
\begin{tabular}{lcccc}
RCC  & \multicolumn{2}{c}{\underline{Sc$^{2+}$}} & \multicolumn{2}{c}{\underline{Y$^{2+}$}} \\
terms & M1 & E2 & M1 & E2 \\
\hline
 & & &  & \\
& \multicolumn{4}{c}{Contributions from effective one-body terms of $\overline{O}$.}\\
 & & \\
$O$                                    & $-$1.55    & $-$1.93    & $-$1.55    & $-$3.54 \\
$\overline{O}-O$                       & 1.78E-2  & 1.37E-2  & 1.30E-2  & 1.98E-2 \\
$\overline{O} S_{1i}$                  & 5.41E-4  & 9.50E-2  & 7.05E-4  & 1.16E-1 \\
$S_{1f}^{\dagger} \overline{O}$        & 3.16E-4  & 9.47E-2  & $-$8.33E-5 & 1.13E-1 \\
$\overline{O} S_{2i}$                  & $-$3.44E-5 & 4.34E-2  & $-$3.35E-4 & 6.99E-2 \\
$S_{2f}^{\dagger} \overline{O} $       & 1.88E-5  & 4.23E-2  & 2.06E-4  & 6.98E-2 \\
$S_{1f}^{\dagger} \overline{O} S_{1i}$ & $-$3.5E-3  & $-$1.14E-2 & $-$2.11E-3 & $-$1.15E-2 \\
$S_{1f}^{\dagger} \overline{O} S_{2i}$ & 1.41E-6  & 3.34E-4  & $-$1.28E-6 & 5.57E-4 \\
$S_{2f}^{\dagger} \overline{O} S_{1i}$ & $-$9.73E-7 & 3.02E-4  & 1.74E-7  & 5.08E-4 \\
$S_{2f}^{\dagger} \overline{O} S_{2i}$ & $-$3.33E-2 & $-$1.92E-2 & $-$2.85E-2 & $-$3.86E-2 \\
 & & & \\
& \multicolumn{4}{c}{Contributions from effective two-body terms of $\overline{O}$.}\\
 & & & \\
$T_1^{\dagger} O S_{2i}$  & 1.57E-3  & 9.66E-4   & 1.38E-3  & 1.73E-3 \\
$S_{2f}^{\dagger} O T_1 $ & 1.57E-3  & 9.65E-4   & 1.37E-3  & 1.72E-3 \\
$T_2^{\dagger} O S_{2i}$  & 0.00     & 0.00      & 5.68E-7  & $-$1.25E-5 \\
$S_{2f}^{\dagger} O T_2 $ & 0.00     & 0.00      & $-$6.29E-7 & $-$1.21E-5 \\
 Norm.                    & 2.28E-2  & 2.44E-2   & 1.99E-2  & 4.06E-2 \\
 & & & \\
 Total                    & $-$1.54 & $-$1.65 & $-$1.54 & $-$3.16  \\
\end{tabular}
\end{ruledtabular}
\label{tab3}
\end{table}

\begin{table}
\caption{Reduced matrix elements of the M1 and E2 transition operators from
$s_{1/2}$ states to the ground states in au in the considered systems. }
\begin{ruledtabular}
\begin{tabular}{lcccc}
RCC & \multicolumn{2}{c}{\underline{Sc$^{2+}$}} & \multicolumn{2}{c}{\underline{Y$^{2+}$}}  \\
terms & M1 & E2 & M1 & E2 \\
\hline
 & & & & \\
& \multicolumn{4}{c}{Contributions from effective one-body terms of $\overline{O}$.}\\
 & & \\
$O$                                    & 3.75E-7   & 4.05     & $-$2.56E-6 & $-$6.70 \\
$\overline{O}-O$                       & $-$8.77E-4  & $-$6.70E-3 & 1.01E-4  & 4.85E-3 \\ 
$\overline{O} S_{1i}$                  & $-$2.61E-5  & $-$3.91E-1 & 7.64E-6  & 4.00E-1 \\
$S_{1f}^{\dagger} \overline{O}$        & $-$2.17E-5  & $-$2.68E-2 & $-$7.74E-6 & 7.83E-2 \\
$\overline{O} S_{2i}$                  & 2.26E-6   & $-$3.25E-2 & $-$1.06E-5 & 5.87E-2 \\
$S_{2f}^{\dagger} \overline{O} $       & 1.72E-6   & 5.74E-3  & $-$1.71E-5 & 1.92E-2 \\
$S_{1f}^{\dagger} \overline{O} S_{1i}$ & $-$7.34E-7  & 1.11E-2  & $-$9.91E-8 & $-$1.93E-2 \\
$S_{1f}^{\dagger} \overline{O} S_{2i}$ & 3.59E-8   & $-$8.21E-4 & $-$2.92E-7 & 1.49E-3 \\
$S_{2f}^{\dagger} \overline{O} S_{1i}$ & 2.93E-8   & 1.90E-3  & $-$1.07E-7 & $-$2.00E-3 \\ 
$S_{2f}^{\dagger} \overline{O} S_{2i}$ & 1.90E-5   & 2.64E-2  & $-$4.09E-4 & $-$5.75E-2 \\
 & & & & \\
& \multicolumn{4}{c}{Contributions from effective two-body terms of $\overline{O}$. }\\
 & & & \\
$T_1^{\dagger} O S_{2i}$   & 1.50E-8  & $-$7.09E-4 & $-$3.58E-8 & 1.60E-3 \\ 
$S_{2f}^{\dagger} O T_1 $  & $-$8.90E-9 & $-$7.74E-4 & 1.55E-8  & 1.63E-3 \\
$T_2^{\dagger} O S_{2i}$   & 0.00     & 0.00     & $-$1.38E-9 & 4.89E-6 \\
$S_{2f}^{\dagger} O T_2 $  & $-$3.01E-7 & 3.45E-3  & 2.24E-7  & $-$4.79E-3 \\ 
 Norm.                     &  1.11E-5 & $-$4.46E-2 & 4.22E-6  & 7.76E-2 \\
 & & & \\
 Total                     & $-$8.92E-4 & 3.59     & $-$3.34E-4 & $-$6.14 \\
\end{tabular}
\end{ruledtabular}
\label{tab4}
\end{table}

\begin{table}
\caption{Reduced matrix elements of the E2 transition operator for the
corresponding $s_{1/2}$ to $d_{5/2}$ states in au in the considered systems.}
\begin{ruledtabular}
\begin{tabular}{lcc}
RCC & Sc$^{2+}$ & Y$^{2+}$  \\
terms  & E2 & E2 \\
\hline
 & & \\
& \multicolumn{2}{c}{Contributions from effective one-body terms of $\overline{O}$.}\\
 & & \\
$O$                                    & 4.97      & $-$8.28 \\
$\overline{O}-O$                       & $-$8.91E-3  & 7.10E-3 \\
$\overline{O} S_{1i}$                  & $-$4.80E-1  & 4.86E-1 \\
$S_{1f}^{\dagger} \overline{O}$        & $-$3.31E-2  & 9.86E-2 \\
$\overline{O} S_{2i}$                  & $-$4.01E-2  & 7.25E-2 \\      
$S_{2f}^{\dagger} \overline{O} $       & 8.80E-3   & 1.75E-2 \\
$S_{1f}^{\dagger} \overline{O} S_{1i}$ & 1.36E-2   & $-$2.36E-2 \\
$S_{1f}^{\dagger} \overline{O} S_{2i}$ & $-$1.01E-3  & 1.82E-3 \\
$S_{2f}^{\dagger} \overline{O} S_{1i}$ & 2.36E-3   & $-$2.50E-3 \\
$S_{2f}^{\dagger} \overline{O} S_{2i}$ & 3.37E-2   & $-$7.25E-2 \\
 & & \\
& \multicolumn{2}{c}{Contributions from effective two-body terms of $\overline{O}$.}\\
 & & \\
$T_1^{\dagger} O S_{2i}$    & $-$8.71E-4  & 1.96E-3 \\
$S_{2f}^{\dagger} O T_1 $   & $-$9.50E-4  & 2.00E-3 \\
$T_2^{\dagger} O S_{2i}$    & 0.00      & 5.59E-6 \\
$S_{2f}^{\dagger} O T_2 $   & 4.23E-3   & $-$5.86E-3 \\
 Norm.                      & $-$5.46E-2  & 9.51E-2 \\
 & & \\
 Total                      & 4.42      & $-$7.60 \\
\end{tabular}
\end{ruledtabular}
\label{tab5}
\end{table}
To understand the importance of the correlation effects, we present the contributions from the individual terms of Eq. (\ref{eqn14}) in Table \ref{tab3} to Table 
\ref{tab5}. The contribution given by the operator $O$ is the Dirac-Fock term and the 
difference between $\overline{O}$ and $O$ represents the pure core-valence (CV) 
correlation effects. The remaining large contributing terms $\overline{O} S_{1i}$ and $\overline{O} S_{2i}$ with their conjugate terms represent the
all-order pair-correlation and core-polarization effects, respectively, for the given valence electron. 

The reduced matrix elements of the M1 and E2 transition operators for the $d_{5/2}\rightarrow d_{3/2}$ transitions in Sc$^{2+}$ and Y$^{2+}$ are given in Table \ref{tab3}. The transition amplitudes due to E2 are slightly larger 
than the amplitudes due to M1 transitions, but the latter totally decide the lifetimes of the 
$d_{5/2}$ states (see Table \ref{tab1}). It is because, the transition probabilities due to E2 transitions depend on fifth power of their wavelengths in the denominator, where as, the M1 transition probabilities depend on third power of their wavelengths in the denominator. The total contribution of the correlation effects, which can be evaluated as the 
difference between the DF and the CCSD(T) results, in the calculation of M1 transition amplitudes are around 0.5\% for both the
systems, however, they are more than 10\% in the case of the E2 transition amplitudes.

The reduced matrix elements of the M1 and E2 transition operators of the transition, $s_{1/2}\rightarrow d_{3/2}$ in Sc$^{2+}$ and Y$^{2+}$ are given in Table \ref{tab4}. The transition amplitudes due to E2 are almost 3-4 orders of magnitude larger than the M1 amplitudes and the former give largest contributions to the lifetimes of the $s_{1/2}$ states. The correlation effects are very large in both the M1 and E2 transitions in Sc$^{2+}$, where as, they are large only in the case of M1 transition in Y$^{2+}$.

In Table \ref{tab5}, we present the E2 transition amplitudes of the $s_{1/2} \rightarrow d_{5/2}$ transitions. The trend of the contributions from different 
RCC terms are similar to the $d_{5/2} \rightarrow s_{1/2}$ transitions of the 
corresponding iso-electronic alkali metal atoms \cite{safronova} and alkaline
earth metal ions \cite{bijaya1}, but these amplitudes are smaller in the considered systems.
The correlation effects in these calculations are just around 12\% and 9\%
in Sc III and Y III, respectively. Again, the correlation effects in M1 transition 
amplitudes of the $d_{5/2} \rightarrow d_{3/2}$ transitions are very small 
compared to the $6p_{3/2} \rightarrow 6p_{1/2}$ transition amplitude in Pb$^+$ \cite{bijaya4}, where we observe similar behavior for the E2 transition amplitudes.

The determination of the lifetimes of the $s_{1/2}$ states depend crucially
on the accurate calculation of the E2 transition amplitudes of both $s_{1/2} \rightarrow d_{3/2}$ 
 and $s_{1/2} \rightarrow d_{5/2}$ transitions, where as, the M1 amplitudes of the former transitions play negligible roles. 

\section{Conclusion}
We have determined the lifetimes of the lowest $d_{5/2}$ and $s_{1/2}$ metastable
states in Sc$^{2+}$ and Y$^{2+}$ which seem to be unavailable in the literature. The transition strengths and the transition probabilities 
of these systems find interesting applications in astrophysics in understanding the post-main sequence evolution of the cool giant stars. These doubly ionized ions may also be suitable for other ion 
trapping experiments as they are reported to have long lifetimes, which need to be explored by the experimentalists.

\section{Acknowledgment}
We are thankful to the authority of C-DAC TeraFlop Supercomputer facilities, 
Bangalore, India for providing their computational resources. One of the authors (BKS) thanks professor Peter Fulde for many useful discussions and for the hospitality at
MPIPKS.


\begin{thebibliography}{26}
\bibitem{rodrigues}
G. C. Rodrigues, M. A. Ourdane, J. Bieron, P. Indelicato and E. Lindroth, Phys. Rev. A {\bf 63}, 012510 (2000)
\bibitem{sergey}
Sergey G. Porsev and Andrei Derevianko, Phys. Rev. A {\bf 73}, 012501 (2006)
\bibitem{bijaya2006}
 Bijaya K. Sahoo, Rajat Chaudhuri, P. P. Das and Debashis Mukherjee, Phys. Rev. Lett. {\bf 96}, 163003 (2006)
\bibitem{das}
B. P. Das, K. V. P. Latha, B. K. Sahoo, C. Sur, R. K. Chaudhuri and D. Mukherjee, J. of Theoretical and computational Chemistry {\bf 4}, 1 (2005)
\bibitem{bijaya1}
B. K. Sahoo, S. Majumder, R. K. Chaudhuri, B. P. Das and D. Mukherjee, J. Phys. B {\bf 37}, 3409 (2004)
\bibitem{bijaya2}
B. K. Sahoo, Md. R. Islam, B. P. Das, R. K. Chaudhuri and D. Mukherjee, Phys. Rev. A {74}, 062504 (2006)
\bibitem{redfors}
A. Redfors, Astron. Astrophys. {\bf 249}, 589 (1991)
\bibitem{guthrie}
B. N. G. Guthrie, Astrophysics and Space Sciences {\bf 10}, 156 (1971)
\bibitem{adelman}
S. J. Adelman, MNRAS {\bf 239}, 487 (1989)
\bibitem{dimitri}
M. S. Dimitrijevic and S. Sahal-Brechot, Journal of Applied Spectroscopy {\bf 65}, 492 (1998) \\
M. S. Dimitrijevic and S. Sahal-Brechot, Joint European and National Meeting (1997)
\bibitem{brage}
Tomas Brage, Glen m. Wahlgren, S. G. Johansson, D. S. Leckrone and C. R. Proffitt, The Astrophysical Journal, {\bf 496}, 1051 (1998) 
\bibitem{maniak}
S. T. Maniak , L. J. Curtis, C. E. Theodosiou, R. Hellborg, S. G. Johansson, I. Martinson, R. E. Irving and D. J. Beideck, Astron. Astrophys. {\bf 286}, 978 (1994)
\bibitem{safronova}
M. S. Safronova, W. R. Johnson and A. Derevianko, Phys. Rev. A {\bf 60}, 4476 (1999)
\bibitem{johnson}
W. R. Johnson, D. R. Plante and J. Sapirstein, Advances in At. Mol. and
Optical physics {\bf 35}, 255 (1995)
\bibitem{mukherjee}
D. Mukherjee, R. Moitra and A. Mukhopadhyay, Mol. Physics {\bf 33}, 955 (1977)
\bibitem{lindgren1}
I. Lindgren, {\it A coupled-cluster approach to the many-body pertubation theory for open-shell systems}, In Per-Olov Löwdin and Yngve Öhrn , editors, Atomic, molecular, and solid-state theory, collision phenomena, and computational methods, International Journal of Quantum Chemistry, Quantum Chemistry Symposium {\bf 12}, pages 33-58, John Wiley \& Sons, March 1978.
\bibitem{bijaya3}
B. K. Sahoo, B. P. Das, R. K. Chaudhuri and D. Mukherjee, Phys. Rev. A {\bf 75}, 032507 (2007)
\bibitem{bijaya4}
B. K. Sahoo, T. Beier, B. P. Das, R. K. Chaudhuri and D. Mukherjee, J. Phys. B {\bf 39}, 355 (2006)
\bibitem{lindgren2}
I. Lindgen and J. Morrison, {\it Atomic Many-Body Theory}, edited by G Ecker, P Lambropoulos, and H Walther ( Springer-Verlag, Berlin, 1985)
\bibitem{nist}
See: http://physics.nist.gov/PhysRefData/Handbook/element$\_$name.htm
\bibitem{epstein1975}
Gabriel L. Epstein and Joseph Reader, J. of Opt. Soc. of America {\bf 65}, 310 (1975)
\end{thebibliography}
\end{document}